\documentclass[11pt]{article}
\usepackage{moriond,epsfig}
\usepackage{multirow}
\usepackage{url}

\def\Journal#1#2#3#4{{#1} {\bf #2}, #3 (#4)}

\def\be{\begin{equation}}
\def\ee{\end{equation}}
\def\bea{\begin{eqnarray}}
\def\eea{\end{eqnarray}}

\begin{document}
\vspace*{4cm}
\title{SEARCH FOR LORENTZ INVARIANCE VIOLATION EFFECTS WITH PKS~2155-304 FLARING PERIOD IN 2006 BY H.E.S.S.}

\author{J. BOLMONT$^1$, R. B\"UHLER$^2$, A. JACHOLKOWSKA$^1$, S.\,J. WAGNER$^3$,\\for the H.E.S.S. COLLABORATION}
\address{$^1$\ LPNHE, Universite Pierre et Marie Curie Paris 6, Universite Denis Diderot Paris 7, CNRS/IN2P3, France, $^2$\ Max-Planck-Instit\"ut f\"ur Kernphysik, Heidelberg, Germany, $^3$\ Landessternwarte, Universit\"at Heidelberg, Germany}

\maketitle\abstracts{
Highly energetic, variable and distant sources such as Active Galactic Nuclei provide a good opportunity to evaluate effects due to the emission and the propagation of high energy photons. In this note, a study of possible energy-dependent time-lags with PKS 2155-304 light curve as measured by H.E.S.S. in July 2006 is presented. These time-lags could either come from the emission processes or also sign a Lorentz Symmetry breaking as predicted in some Quantum Gravity models. A Cross-Correlation function and a Wavelet Transform were used to measure the time-lags. The 95\% Confidence Limit on the Quantum Gravity energy scale based on the statistical and systematic error evaluation was found to be $7\times10^{17}$ GeV considering a linear correction in the standard photon dispersion relations and assuming that emission-induced time-lags are negligible. For now, this limit is the best ever obtained with a blazar.}

\section{Introduction}

Quantum Gravity phenomenology \cite{amelino08} has known a growing interest in the past decade, especially since the time 
it was argued that the quantum nature of space-time could have measurable effects on photon propagation over large distances. Namely, the quantum nature of the space-time at the Planck scale could induce a dependance of the light group velocity with the energy of the photons \cite{amelino98}$^,$\cite{ellis08_mattingly}.

It is generally assumed that the light velocity is modified following to
\begin{equation}
c' = c ~ \left( 1 +  \xi \frac{E}{\mathrm{E}_{\rm P}} + \zeta \frac{E^2}{
  \mathrm{E}^2_{\rm P}}\right)
\label{eq:correction}
\end{equation}
at the second order, where $\mathrm{E}_{\rm P} = 1.22 \times 10^{19}$~GeV is the Planck energy and where $\xi$ and $\zeta$ are free parameters which need to be determined. In the following, only the linear correction will be considered and a limit on $E_{\rm QG} = |\xi|^{-1} \mathrm{E}_{\rm P}$ will be set.

As $\mathrm{E}_{\rm P}$ is very large, the modification of the velocity of the photons is expected to be tiny. However, as it is related to the nature of the space-time, it was proposed \cite{amelino98} that these tiny effects could add up over long distances and lead to detectable time-lags between photons of different energies, assuming these photons were emitted \textit{at the same time}. Taking into account the expansion of the Universe \cite{jacob}, the time-lag ${\rm\Delta} t$ is obtained from the dispersion relations (\ref{eq:correction}):
\begin{equation}
\label{eq:linear}
 \frac{{\rm\Delta} t}{{\rm\Delta} E} \approx \frac{\xi}{\mathrm{E}_{\rm P} \mathrm{H}_0} \int_0^z dz'
   \frac{(1+z')}{\sqrt{\Omega_m(1+z')^3 + \Omega_{\Lambda}}},
\end{equation}
in the case of the linear correction and where $\Omega_m$, $\Omega_{\Lambda}$ and $\mathrm{H}_0$ are parameters of the Cosmological 
Standard Model ($\Omega_m$ = 0.3, $\Omega_{\Lambda}$ = 0.7 and $\mathrm{H}_0$ = 70~km s$^{-1}$ Mpc$^{-1}$). ${\rm\Delta}E$ is 
the energy difference of the photons. The time-lag may decrease or increase with ${\rm\Delta}E$ depending on the model in consideration. The integral term increases with the redshift and takes into account the fact that two photons travelling with two different speeds do not take the same time to cross the Universe and then do not see the same expansion. 

The goal of the 'time of flight' studies is to measure the time-lags, which should be energy dependant. For this, 
two kind of variable extragalactic sources may be considered: the Gamma-Ray Bursts (GRBs) and the Active Galactic 
Nuclei (AGNs). However, both these sources could introduce intrinsic time lags in the measurements. Therefore, it would be 
necessary to study the possible effects as a function of the redshift of the source. When this is not possible, the intrinsic (or source) effects are often assumed to be negligible. In the present study, this hypothesis was adopted.

\begin{figure}[t!]
   \centering
   \includegraphics[width=0.5\textwidth]{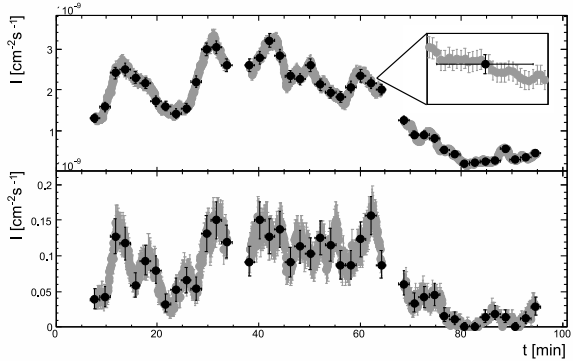} 
   \caption{Light curve of the PKS2155-304 flare during the night of July 28 in 2006. Top: 200-800 GeV. Bottom: $>$~800 GeV. The original data points (in black) are binned in two-minute time intervals. The zero time point is set to MJD 53944.02. Gray points show the oversampled light curve, for which the two-minute bins are shifted in units of five seconds.}
   \label{fig:1}
\end{figure}

\section{Data and methods}

In this note (more details are available elsewhere \cite{aharonian08}), the search for propagation effects with the blazar PKS 2155-304 is described. During the night of the 28th of July 2006, the H.E.S.S. experiment detected \cite{BIGFLARE} an exceptional flare of this source, located at z~=~0.116, with a high flux (10 000 photons recorded in 1.5~hours) and a high variability (rise and fall times of $\sim$200~s). The over-sampled light curve of the flare is shown in Fig.~\ref{fig:1} in two different energy bands.

\begin{figure}[t!] 
   \centering
   \includegraphics[width=0.6\textwidth]{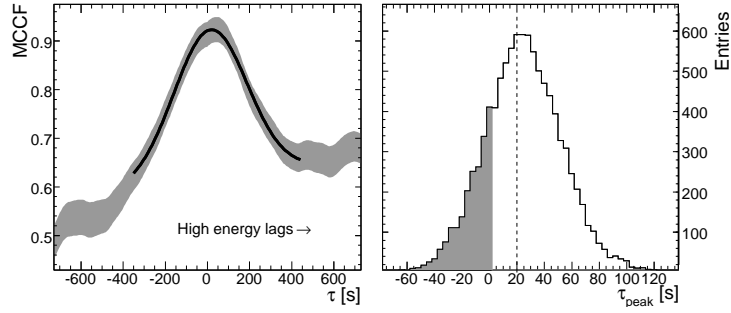} 
   \caption{Left: the MCCF obtained with the two light curves of Fig.~\ref{fig:1}. Right: the MCCF peak distribution (CCPD) obtained with 10000 realizations of the light curves.}
   \label{fig:2}
\end{figure}

To measure the time lags between photons in two different light curves, two independant analyses were carried out, using 
two different methods:
\begin{itemize}
\item The position of the maximum of the Modified Cross Correlation Function \cite{li} (MCCF) gives directly the value of the time lag. This method was applied to the oversampled light curves of Fig.~\ref{fig:1} in the energy bands 200-800 GeV and $>$~800 GeV (Fig.~\ref{fig:2}, left) which corresponds to $<\Delta E>\ \sim$1~TeV;
\item Following Ellis {\it et al.} \cite{ellis04}, the Continuous Wavelet Transform \cite{mallat} (CWT) was used to locate the extrema of the light curves with great precision. An extremum of the low energy band was associated with an extremum in the high energy band to form a pair. Selection criteria were applied to reject fake extrema. The energy bands 210-250 GeV and $>$~600 GeV were used, with a bin width of one minute and no oversampling of the light curves. These energy bands give a mean $\Delta E$ of $\sim$0.92~TeV.
\end{itemize}

\begin{figure}[t!] 
   \begin{minipage}[b]{0.45\textwidth}
   \centering
   \includegraphics[width=\textwidth]{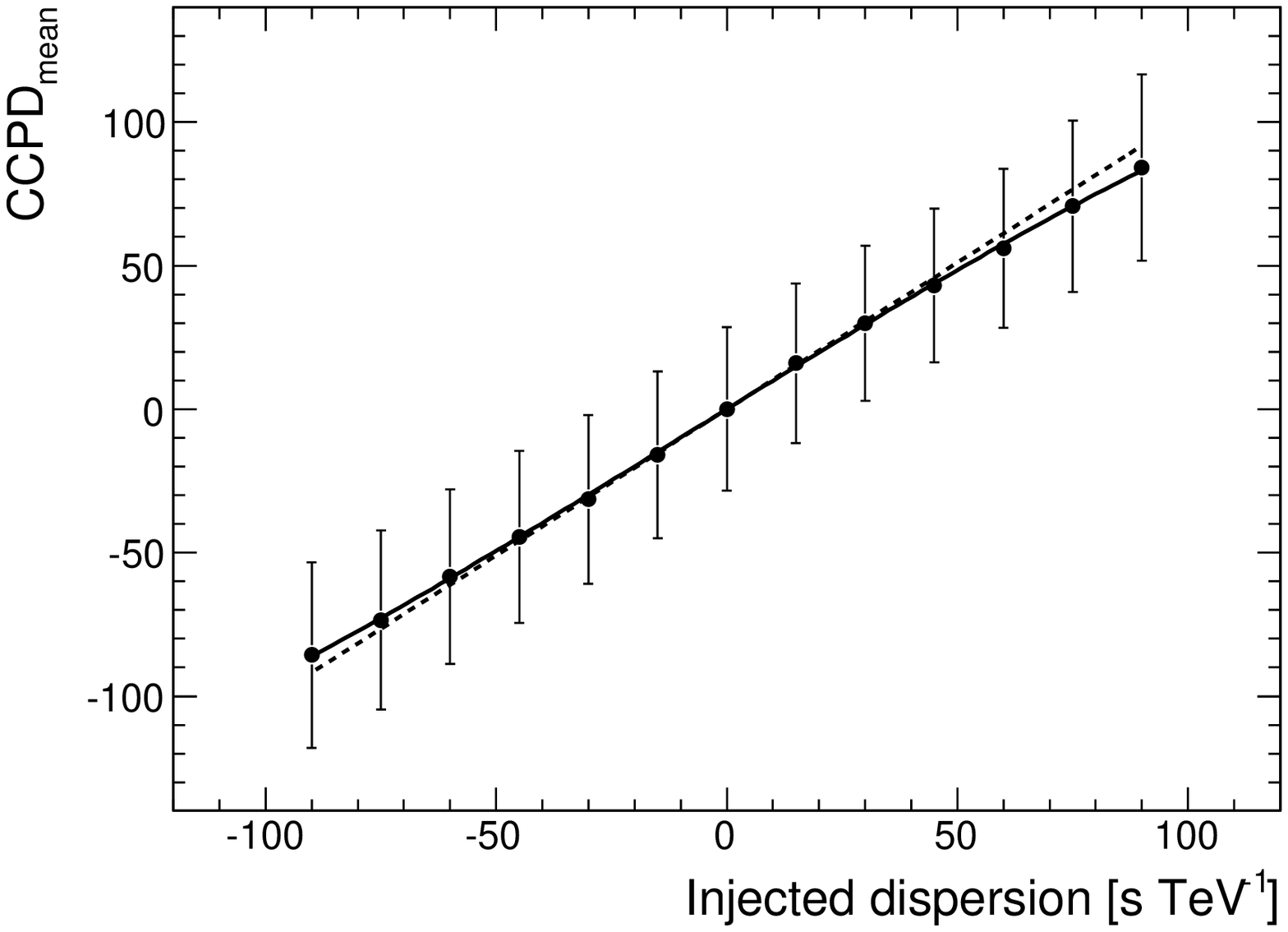} 
   \caption{Mean of the MCCF peak distribution (CCPD) as a function of the injected dispersion. Each CCPD is obtained from ten thousand simulated light curves. The points have been shifted by the mean value of the CCPD of the original data (Fig.~\ref{fig:2}). The dashed line shows the linear response function.}
   \label{fig:3}
   \end{minipage}\hfill
   \begin{minipage}[b]{0.45\textwidth}
   \centering
   \includegraphics[width=0.9\textwidth]{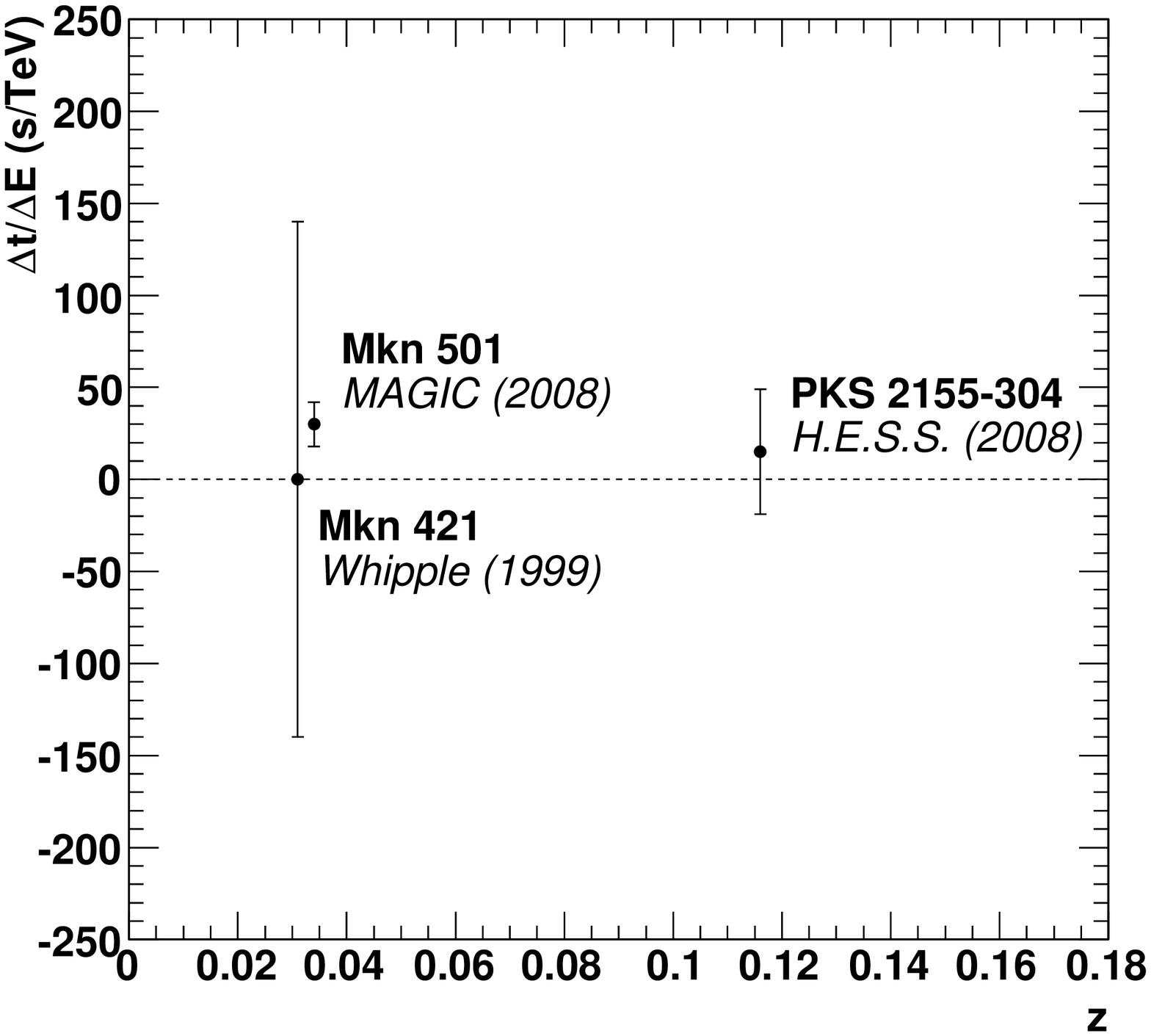} 
   \caption{The dispersion measured as a function of the redshift. The results of the Whipple and the MAGIC experiments are compared with the present H.E.S.S. result.}
   \label{fig:4}
   \end{minipage}
\end{figure}

As shown in Fig.~\ref{fig:2} (left), the MCCF was fitted with the sum of a gaussian and a first degree polynomial to determine the position of the maximum. It gave $\tau_{\rm peak}$ = 20~s. In order to evaluate the uncertainties on this result, 10000 
light curves were simulated in each energy band varying the flux within the error bars. For each pair of light curves, 
the MCCF was computed and its maximum was filled into the distribution shown in Fig.~\ref{fig:2} (right). This distribution 
is slightly asymmetric, with a mean of 25~s and an RMS of 28~s. Another test, performed by injecting a dispersion in the 
data (Fig.~Ê\ref{fig:3}), showed no significant deviation of the measured lag. Then, as the time-lag obtained was compatible 
with zero, a 95\% confidence upper limit on the linear dispersion of 73~s/TeV was found.

With the CWT method, two pairs of extrema were obtained giving a mean time delay of 27 seconds. A method similar to 
the one used for the CCF was applied to determine the errors, which were found to be in a range between 30 and 36~seconds. 
A 95$\%$ confidence limit of 100~s/TeV was obtained for the linear correction.

\begin{table}[t]
\caption{The results obtained with the two methods in the case of a linear dispersion in energy.}
\vspace{0.4cm}
\begin{center}
\begin{tabular}{|c|c|c|c|c|}
\hline
Method                 & Energy bands              & $<{\rm\Delta} E>$              & $({\rm\Delta} t/{\rm\Delta} E)_{95\%\ \mathrm{CL}}$ & $E_{\mathrm{QG}\ 95\%\ \mathrm{CL}}$ \\
\hline
\multirow{2}{*}{MCCF}  & 200 GeV $<$ E $<$ 800 GeV & \multirow{2}{*}{1.02 TeV} & \multirow{2}{*}{$<$ 73 s/TeV}             & \multirow{2}{*}{$>$ $7.2\times10^{17}$ GeV}\\
                       & and E $>$ 800 GeV         &                           &                                           &                                            \\
\hline
\multirow{2}{*}{CWT}   & 210 GeV $<$ E $<$ 250 GeV & \multirow{2}{*}{0.92 TeV} & \multirow{2}{*}{$<$ 100 s/TeV}            & \multirow{2}{*}{$>$ $5.2\times10^{17}$ GeV}\\
                       & and E $>$ 600 GeV         &                           &                                           &                                            \\
\hline
\end{tabular}
\end{center}
\label{tab:res}
\end{table}%

\section{Results and discussion}

The results obtained with the two methods are summarized in Table~\ref{tab:res} (see next page). The MCCF leads to a limit of $E_\mathrm{QG} > 7.2\times10^{17}$~GeV. The CWT gives a lower limit of $E_\mathrm{QG} > 5.2\times10^{17}$ GeV, mainly due to a larger measured time-lag and a lower value of $<\Delta E>$. 

Fig.~\ref{fig:4} (previous page) shows all the results obtained so far with AGNs. The Whipple collaboration set a limit of 
$4\times10^{16}$ GeV using a flare of Mkn~421 (z = 0.031) in 1996 \cite{whipple}. More recently, the MAGIC collaboration 
obtained a limit of $2.6\times10^{17}$ GeV with a flare of Mkn~501 (z = 0.034) \cite{magic}. The results obtained with 
H.E.S.S. are more constraining due to the fact that $(i)$ PKS~2155-304 is almost four times more distant than Mkn~421 and Mkn~501, $(ii)$ the statistics are higher by a factor of ten. 

As mentionned above, GRBs are other good candidates for time of flight studies. Population studies have already been 
carried out \cite{grbs} and lead to limits of the order of $10^{16}$ GeV. The best limit so far is 
$E_\mathrm{QG} > 2\times10^{18}$ GeV. It has been obtained by the Fermi experiment \cite{fermi} with GRB~080916C 
(z~=~4.35) using both GBM and LAT data. However, the time lag was obtained comparing the arrival time of the highest energy photon (13.2~GeV) with the trigger time, leading to $\Delta t = 16.5$~s. A more detailed analysis including error calibration would be necessary.

As far as the search for the Quantum Gravity propagation effect is concerned, the GRBs and the AGNs are complementary. 
AGN flares can be detected with high statistics with ground-based gamma-ray telescopes, which give large values of 
$\Delta E \sim 1$ TeV. However, the absorption of the high energy photons by the extra-galactic background light limits 
the distance of the observed objects. On the other hand, GRBs are easilly detected by satellite experiments at very 
high redshifts (up to $\sim$6) and up to a few hundreds of GeV ($\Delta E \sim 10$ GeV). Till now, no significant result has been obtained with either AGNs or GRBs for the linear and quadratic terms in the photon dispersion relations. Further observations of both a high number of AGN flares and GRBs will be necessary to give robust conclusions on possible propagation effects. Present and future experiments such as Fermi, CTA or AGIS~\cite{ctaagis} will greatly improve our capabilities in this area.

\end{document}